\documentclass[
    ,final            
  ]
  {aipproc}

\layoutstyle{6x9}
\newcommand{\lesssim}{\lower.5ex\hbox{$\; \buildrel < \over\sim \;$}}
\newcommand{\gtrsim}{\lower.5ex\hbox{$\; \buildrel > \over\sim \;$}}

\newcommand{\apj}{{\it Astrophys. J.}}

\begin{document}

\title{Neutrinos and Gamma Rays from Photomeson Processes in 
Gamma Ray Bursts}

\author{Armen Atoyan}{
  address={Centre de Recherche Math\'ematiques, 
Universit\'e de Montr\'eal, Montr\'eal, Canada H3C 3J7}
}

\author{Charles D.\ Dermer}{
  address={Code 7653, Naval Research Laboratory, 
Washington, DC 20375-5352 USA}
}

\begin{abstract}
Acceleration of high-energy hadrons in GRB blast waves will be
established if high-energy neutrinos are detected from GRBs.  Recent
calculations of photomeson neutrino production are reviewed, and new
calculations of high-energy neutrinos and the accompanying hadronic
cascade radiation are presented. If hadrons are injected in GRB blast
waves with an energy corresponding to the measured hard X-ray/soft
$\gamma$-ray emission, then only the most powerful bursts at fluence
levels $\gtrsim 3\times 10^{-4}\,\rm erg \, cm^{-2}$ offer a realistic
prospect for detection of $\nu_\mu$.  Detection of high-energy
neutrinos are likely if GRB blast waves have large baryon loads and
Doppler factors $\lesssim 200$. Significant limitations on the
hadronic baryon loading and the number of expected neutrinos are
imposed by the fluxes from pair-photon cascades initiated in the same
processes that produce neutrinos.
\end{abstract}

\maketitle


\section{Introduction}

In recent work \cite{da03}, we considered neutrino production for two
leading scenarios for the sources that power long-duration GRBs,
namely the collapsar \cite{woo} and supranova (SA)
\cite{vs} models. In the collapsar model,  the core of a 
massive star collapses directly to a black hole, and the most
important radiation field for photomeson neutrino production is
 the internal synchrotron radiation field
\cite{wb}.  In the supranova (SA) model, a pulsar nebula synchrotron
radiation within an expanding supernova remnant (SNR) shell
\cite{kg02} provides additional external photon target for photomeson
interactions (see also Refs.\ \cite{rmw03,gg03}).

We found that the presence of the external field in the SA model can
increase the number of detectable neutrinos by an order of magnitude
or more over the collapsar model when the Doppler factor $\delta
\gtrsim 200$. When $\delta \lesssim 200$, the internal synchrotron
field can become effective for photomeson interactions. In our
calculations, we assumed that the energy injected in protons is equal
to the energy of electrons producing the photon fluence measured at
X-ray and $\gamma$-ray energies. In both  models, the likelihood of
detecting a neutrino from a GRB with a km-scale detector such as
IceCube is small except for rare GRBs with fluence $ \gtrsim 3\times
10^{-4}\,\rm erg \, cm^{-2}$. If GRB blast waves are strongly
baryon-loaded, however, as required in a GRB model for high-energy
cosmic rays \cite{wda03}, then we predict that 100 TeV -- 100 PeV
neutrinos will be detected several times per year with IceCube when
$\delta \lesssim 200$.

Here we summarize our calculations of photomeson neutrino production
for the collapsar and SA models, and present new calculations for the
hadronic cascade radiation from a GRB.

\section{Model}

The GRB model is adapted from our photo-hadronic model for blazar jets
\cite{ad01}, and takes into account the injection of 
nonthermal protons which lose energy through photomeson interactions.
Protons are injected with a number spectrum $\propto E_p^{-2}$ at
comoving proton energies $E_p > \Gamma $ GeV up to a
maximum proton energy determined by the condition that the particle
Larmor radius is smaller than both the size scale of the emitting
region and the photomeson energy-loss length. Here $\Gamma$ is the
Lorentz factor of the blast wave.  The observed synchrotron spectral
flux in the prompt phase of the burst is parameterized by the
expression $F(\nu) \propto
\nu^{-1} (\nu/\nu_{br})^{\alpha}$, where $h\nu_{br}=300$ keV,
$\alpha = -0.5$ above $\nu_{br}$ and an exponential cutoff
at 10 MeV, and $\alpha = 0.5$ when $10\, {\rm
keV} \leq h\nu \leq h \nu_{br}$. At lower energies, $\alpha= 4/3$.
The observed total hard X-ray/soft $\gamma$-ray photon fluence $
\Phi_{tot} \cong t_{dur}\int_0^\infty d\nu F(\nu )$, where $t_{dur}$
is the characteristic duration of the GRB. We consider a source at
redshift $z=1$ and take the hard X-ray/soft $\gamma$-ray fluence
$\Phi_{tot} \gtrsim 3\times 10^{-5} \,\rm erg \; cm^{-2}$.  Two or
three GRBs should occur each month above this fluence level.

A total amount of energy $E^\prime = 4\pi d_L^2\Phi_{tot} \delta^{-3}
(1+z)^{-1}$ is injected in the form of accelerated proton energy into
the comoving frame of the GRB blast wave. Here $z$ is the redshift and
$d_L$ is the luminosity distance. The energy deposited into each of
$N_{sp}$ light-curve pulses (or spikes) is therefore $E_{sp}^\prime =
E^\prime/N_{sp}\,$ ergs.  We assume that all the energy
$E_{sp}^\prime$ is injected in the first half of the time interval of
the pulse (to ensure variability in the GRB light curve), which
effectively corresponds to a characteristic variability time scale
$t_{var} = t_{dur}/2N_{sp}$.  The proper width of the radiating region
forming the pulse is $\Delta R^\prime \cong t_{var} c\delta/(1+z)$,
from which the energy density of the synchrotron radiation can be
determined \cite{ad01}.  We set the GRB prompt duration $t_{dur} =
100\,$s, and let $N_{sp} = 50$, corresponding to $t_{var} = 1\,\rm s$.
The magnetic field is determined by assuming equipartition between the
energy densities of the magnetic field and the electron energy.

For the SA model, we assume the existence of an external radiation
field given by the expression $\nu L_\nu \propto\nu^{1/2}
\exp(-\nu/\nu_{ext})$, with $ h\nu_{ext}
\approx 1$ keV \cite{kg02}.  The intensity
of this field is determined by the assumption that the integral power
$L_{ext} =\int_0^\infty L_\nu \rm d \nu$ is equal to the power of the
pulsar wind $L_{pw}\approx (10^{53}\,{\rm erg})/t_{delay}$, assuming
that a total of $\approx 10^{53} \,\rm erg$ of pulsar rotation energy
is radiated during the time $t_{delay}$ (which is here set equal to
0.1 yr) from the rotating supramassive neutron star before it
collapses to a black hole.  The energy $h\nu_{ext}\simeq 0.1
\,\rm keV$ is the characteristic energy of synchrotron radiation
emitted by electrons (of the pulsar wind) with Lorentz factors
$\gamma_{pw}\sim 3\times 10^4$ in a randomly ordered magnetic field of
strength $\approx 10$ G. The radius $R = 0.05ct$ is determined by
assuming that $0.05c$ is the mean speed of the SNR shell, and that the
external photon energy density $\propto L/2\pi R^2$.

Fig.\ 1 shows the total $\nu_\mu$ fluences expected from a model GRB
with $N_{sp}=50$ pulses.  The thin curves show collapsar model results
at $\delta = 100,$ 200, and 300, with $\Gamma = \delta$. The
expected numbers of $\nu_\mu$ that a km-scale detector such as IceCube
would detect are $N_\nu = 3.2\times 10^{-3}$, $1.5\times 10^{-4}$, and
$1.9\times 10^{-5}$, respectively.  There is no prospect to detect
$\nu_\mu$ from GRBs at these levels.  The heavy solid and dashed
curves in Fig.\ 1 give the SA model predictions of $N_\nu = 0.009$ for
both $\delta = 100$ and $\delta = 300$. The equipartition magnetic
fields are 1.9 kG and 0.25 kG, respectively. The external radiation
field in the SA model makes the neutrino detection rate insensitive to
the value of $\delta$ (as well as to $t_{var} \gtrsim 0.1\,\rm s$, as
verified by calculations), but there is still little hope that a
km$^3$ detector could detect such GRBs. Neutrino production efficiency
would improve by at most a factor of 3 in the collapsar model if
$t_{var} \sim 1\,$ms (and $N_{sp} = 5\times 10^4$ to provide the same
total fluence). Such narrow spikes are, however, then nearly opaque to
$\gamma$ rays with energies $\gtrsim 100$ MeV \cite{da03}. Even in this
case, there is little hope to detect neutrinos except from rare GRBs
with fluence $ \Phi_{tot} \gtrsim 3\times 10^{-4}\,\rm erg \,
cm^{-2}$.

\begin{figure}[t]
\vspace*{15.0mm} %
\includegraphics[width=10.cm]{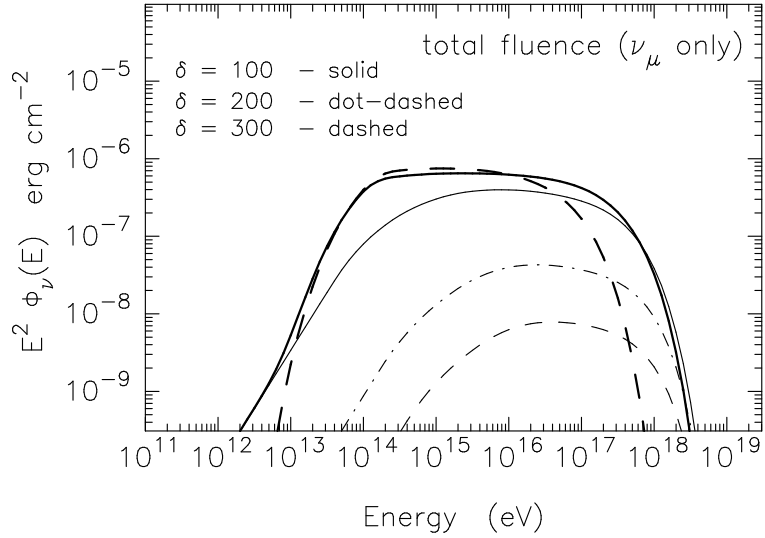}
\caption{Energy fluence of photomeson muon neutrinos ($\nu_\mu)$ for 
a model GRB with hard X-ray fluence $\Phi_{tot} = 3\times 10^{-5}
\,\rm erg \; cm^{-2}$ for different Doppler factors $\delta$. 
The thin curves show collapsar model results
where only the internal synchrotron radiation field provides a source
of target photons. }
\label{fig1}
\end{figure}

Fig.\ 2 shows new calculations of photomeson neutrino production for a
GRB with $ \Phi_{tot} = 3\times 10^{-4}\,\rm erg \, cm^{-2}$ and
$\delta = 100$, as well as the accompanying electromagnetic radiation
induced by pair-photon cascades from the secondary electrons and
$\gamma$ rays from the same photomeson interactions. The total number
of $\nu_\mu$ expected with IceCube is $\cong 0.1$. The total fluence
of cascade photons shown here is contributed by lepton synchrotron
(dot-dashed) and Compton (dashed) emissions. For comparison, the
dotted curve shows the primary lepton synchrotron radiation spectrum
assumed for the calculations. The level of the fluence of the cascade
photons is $\approx 10$\% of the primary synchrotron radiation. This
means that the maximum allowed baryon loading for these parameters
cannot exceed a factor of $\approx 30$ in order not to overproduce the
primary synchrotron radiation fluence. This limits the maximum number
of $\nu_\mu$ to $\approx 3$ even in the case of large baryon
loading for rare, powerful GRBs. These numbers cannot be further
increased in the SA model because of the efficient extraction that is
already provided by internal photons alone for these parameters.

\begin{figure}[t]
\vspace*{15.0mm} %
\includegraphics[width=11.0cm]{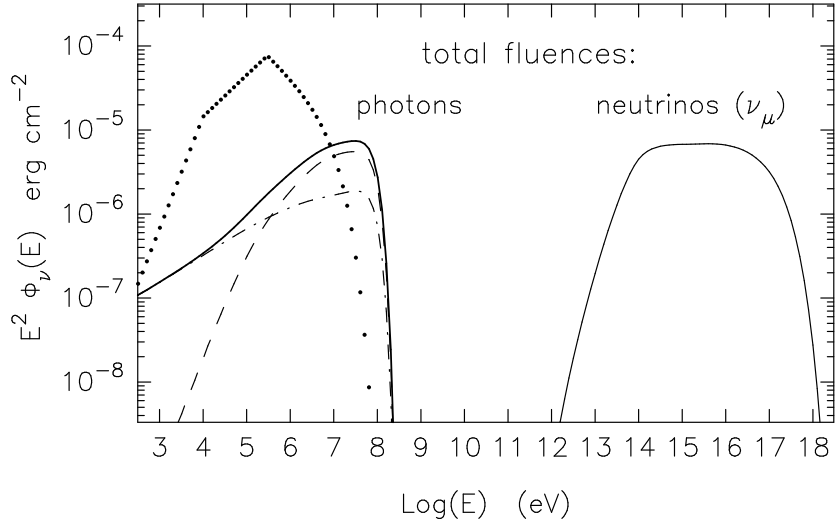}
\caption{Energy fluence of photons and photomeson muon neutrinos for 
a collapsar-model GRB with hard X-ray fluence $\Phi_{tot} = 3\times 10^{-4}
\,\rm erg \; cm^{-2}$ and $\delta = 100$. The dotted curve shows the 
fluence of a GRB used for calculations, and the dashed and dot-dashed
curves show the Compton and synchrotron contributions to the photon
fluence from the electromagnetic cascade initiated by secondaries from 
 photomeson processes, respectively.  }
\label{fig2}
\end{figure}

In conclusion, we predict that at most a few high-energy
$\nu_\mu$ can be detected with IceCube even from a very bright
GRB at the fluence level $\Phi_{tot} \gtrsim 3\times 10^{-4}
\,\rm erg \; cm^{-2}$, and only when the baryon loading is 
high \cite{da03,wda03}. This is because the detection of a single $\nu_\mu$
requires a $\nu_\mu$ fluence $ \gtrsim 10^{-4}
\,\rm erg \; cm^{-2}$ above 1 TeV. Since the energy release in high-energy
neutrinos and electromagnetic secondaries is about equal, this energy
will be reprocessed in the pair-photon cascade and emerge in the form
of observable radiation at hard X-ray/soft $\gamma$-ray energies, and
this radiation cannot exceed the measured fluence in this regime. This
imposes a robust limit on the maximum number of $\nu_\mu$ even from a 
GRB with
very high baryon loading.

\begin{theacknowledgments}
AA thanks the NRL High Energy Space Environment Branch for support
and hospitality during visits. The work of CD is supported by the
Office of Naval Research and NASA GLAST science investigation grant
DPR \# S-15634-Y.
\end{theacknowledgments}


\begin{thebibliography}{}

\bibitem{da03} C.\ D.\ Dermer and A.\ M.\ Atoyan, {\it Phys.\ Rev.\ Lett.}
{\bf ~91}, 071102, (2003).

\bibitem{woo} C.\ L.\ Fryer, S. \ E.\
Woosley, and D.\ H.\ Hartmann \apj {\bf ~526}, 152 (1999).

\bibitem{vs} M.\ Vietri and L.\ Stella, \apj {\bf~ 507}, L45 (1998).

\bibitem{wb}E.\ Waxman and J.\ N.\ Bahcall, {\it Phys.\ Rev.\ Lett.}
 {\bf ~78}, 2292 (1997).


\bibitem{kg02} A.\ K{\" o}nigl and J.\ Granot, \apj {\bf ~574}, 134 (2002).

\bibitem{rmw03} Razzaque, S., Meszaros, P., and Waxman, E.\ 
 {\it Phys.\ Rev.\ Lett.} {\bf ~90}, 241103 (2003).

\bibitem{gg03} Guetta, D.~\& Granot,
J.\ {\it Phys.\ Rev.\ Lett.}, {\bf 90}, 201103 (2003). 

\bibitem{wda03} S.\ D.\ Wick, C.\ D. Dermer, and A.\ Atoyan, 
{\it Astroparticle Physics}, submitted (astro-ph/0310667).


\bibitem{ad01} A.\ M.\ Atoyan and C.\ D.\ Dermer, \apj
{\bf ~586}, 79, (2003).




\end{thebibliography}
\end{document}